%% file: BlackFlower.tex
\documentclass[11pt,a4paper]{article}
\usepackage{jheppub}

\usepackage[dvipsnames]{xcolor}
\usepackage{hyperref}
\hypersetup{
 colorlinks,
 citecolor=blue,
 linkcolor=magenta,
 urlcolor=blue}
\usepackage[T1]{fontenc}
\usepackage[utf8]{inputenc}
\usepackage{graphicx}
\graphicspath{ {./images/} }
\usepackage{amsmath}
\usepackage{amssymb}
\usepackage{mathtools}
\usepackage{braket}
\usepackage{float}
\usepackage{cancel}
\DeclareMathAlphabet{\mathbbold}{U}{bbold}{m}{n}
\usepackage{parskip}
\usepackage{array}
\usepackage{xfrac}
\numberwithin{equation}{section}
\usepackage{setspace}
\input{gdefn.tex}

\title{{\bfseries{Black Flower Microstates}}}
\author[a]{Suvankar Dutta}
\author[a]{Shruti Menon}
\affiliation[a]{\small Department of Physics, Indian Institute of Science Education and Research Bhopal, Bhopal Bypass Road, Bhopal - 462066, India}

\emailAdd{suvankar@iiserb.ac.in}
\emailAdd{shruti98.menon@gmail.com}

\abstract{We investigate stationary, non-axisymmetric black hole solutions in AdS$_3$ gravity, known as black flower geometries, in the Chern--Simons formulation. Boundary conditions are specified by a collective field theory--inspired Hamiltonian with field-dependent chemical potentials and angularly inhomogeneous boundary data. We construct a tractable class of solutions and analyze their geometric and thermodynamic properties, obtaining an entropy with nontrivial dependence on the angular deformation. Upon quantization of the boundary theory via bosonization, the boundary degrees of freedom are mapped to relativistic free fermions. We explicitly construct and count the microstates associated with a given black flower geometry and find exact agreement with the Bekenstein--Hawking entropy.

}

\begin{document}
\maketitle
\flushbottom

\section{Introduction}

The Ba\~nados--Teitelboim--Zanelli (BTZ) black hole provides a genuine black hole solution in three-dimensional $\mathrm{AdS}_3$ gravity, despite the absence of local bulk degrees of freedom \cite{banados:1992,Banados:1998gg}. Its entropy admits a precise microscopic interpretation via asymptotic symmetries and the Cardy formula, making it a paradigmatic setting for holographic studies of black hole thermodynamics \cite{Strominger:1997eq,Carlip:1998qw,Carlip:2000nv}.

More recently, attention has shifted to less symmetric $\mathrm{AdS}_3$ black hole solutions, such as the stationary but generically non-axisymmetric \emph{black flower} geometries \cite{Afshar:2016wfy,Grumiller:2016kcp,Afshar:2016kjj,Sammani:2024ugh}. In these spacetimes the horizon develops angular modulations while remaining regular and asymptotically AdS, providing a natural arena to explore how black hole microstate counting extends beyond rotationally symmetric settings. 

A powerful framework for studying such geometries is the Chern--Simons formulation of three-dimensional gravity with a negative cosmological constant. In this description, gravity is equivalent to two decoupled \(SL(2,\mathbb{R})\) Chern--Simons theories, and the bulk dynamics is entirely encoded in boundary data \cite{Achucarro:1986uwr,Witten:1988hc}. Different choices of boundary conditions correspond to different boundary Hamiltonians, which determine the dynamics of boundary gravitons. Importantly, one is not restricted to the standard Brown--Henneaux boundary conditions \cite{Brown:1986nw}, and a number of alternative boundary conditions have been extensively studied in the literature~\cite{Henneaux:2013dra, Bunster:2014mua, Perez:2016vqo, Perez:2012cf, Afshar:2016wfy, Grumiller:2016pqb, Grumiller:2016kcp, Grumiller:2019tyl, Afshar:2016kjj, Afshar:2013vka, Troessaert:2013fma, Avery:2013dja, Ammon:2017vwt, Ozer:2019nkv, Gonzalez:2018jgp, Campoleoni:2010zq,Ojeda:2020bgz}. More general choices allow the chemical potentials---the temporal components of the gauge fields---to depend explicitly on the dynamical boundary fields, with their functional form fixed by the chosen boundary Hamiltonian.

In this work, we employ the \emph{collective field theory} (ColFT) boundary Hamiltonian of Jevicki and Sakita \cite{jevicki,sakita}, which arises from unitary matrix quantum mechanics and describes an effective one-dimensional fluid on the boundary. Within this framework, black flower geometries appear as time-independent, non-axisymmetric bulk solutions corresponding to nontrivial boundary fluid configurations.

The main aim of this paper is to analyze black flower geometries within the ColFT boundary condition in the presence of an arbitrary potential. In the absence of the potential, the system preserves spherical symmetry and its microstates were identified previously \cite{Sheikh-Jabbari:2016npa,Afshar:2016uax,Afshar:2017okz,Dutta:2025uch, Dutta:2025ypr}. Turning on the potential breaks this symmetry and alters the relation between conserved charges and horizon geometry, leading to a nontrivial potential dependence of the black flower entropy.

We quantize the ColFT in the presence of the potential and construct the corresponding Hilbert space in a systematic manner. Promoting the classical Poisson brackets to quantum commutators leads to a \(U(1)\) Kac--Moody current algebra without a central extension. Using bosonization, the collective fields are represented in terms of relativistic free fermions, allowing bulk geometries to be identified with bosonic excitations built from fermionic particle--hole states. In this fermionic formulation, black flower microstates with fixed mass and angular momentum arise as excitations above suitable ground states.

When the potential is included, the microscopic counting problem becomes more involved and is naturally analyzed perturbatively in the coupling \(\lambda\), which controls the strength of the potential. Within this perturbative framework, we explicitly construct the relevant microstates and compute their degeneracy. The resulting microscopic entropy reproduces the Bekenstein--Hawking entropy of the black flower geometry, including its characteristic dependence on the potential.

Taken together, our results demonstrate that even in the presence of symmetry-breaking deformations induced by an arbitrary boundary potential, black hole entropy continues to admit a consistent and precise microscopic interpretation.

The paper is organized as follows. Section~2 reviews AdS$_3$ gravity in the Chern--Simons formulation and introduces boundary Hamiltonians, with emphasis on the collective field theory choice. In Section~3, we construct stationary, non-axisymmetric black flower solutions and analyze their geometry and thermodynamics. Section~4 presents the microscopic quantization and state counting that reproduces the black flower entropy, and Section~5 contains a summary and outlook.

\section{A Brief Review of AdS\(_3\) Gravity, Chern--Simons Theory, and Collective Field Theory Boundary Conditions}
\label{sec:Ads3gravity}

This section briefly reviews the formulation of three-dimensional Anti--de Sitter (AdS$_3$) gravity as a Chern--Simons gauge theory, emphasizing the role of boundary Hamiltonians. The bulk theory is topological, with all nontrivial dynamics encoded in boundary degrees of freedom, and different choices of boundary Hamiltonian define distinct boundary conditions and boundary dynamics. In this work, we focus on the ColFT Hamiltonian. Further details may be found in \cite{Dutta:2023uxe, Dutta:2025uch}.

Gravity in $\mathrm{AdS}_{3}$ can be described as a Chern--Simons theory with gauge group 
$SO(2,2)\simeq SL(2,\mathbb{R})\oplus SL(2,\mathbb{R})$ \cite{Campoleoni:2010zq,Campoleoni:2011hg}. 
The gauge fields split into two chiral sectors $A^\pm$, related to the vielbein and spin connection by
\begin{equation}
A^\pm=\left(\omega^a\pm \frac{1}{l}e^a\right)T_a,
\end{equation}
so that the Einstein--Hilbert action becomes the difference of two Chern--Simons actions with level $\mathrm{k}$
\begin{align}\label{eq:CSacn}
I = I_{CS}(A^+) - I_{CS}(A^-)
\end{align}
where the Chern-Simons action for \(A^\pm\) is given by
\begin{align}\label{eq:CSacn2}
I_{CS}(A^\pm) = \frac{\mathrm{k}}{4\pi} \int \Tr\left[ A^\pm \wedge dA^\pm + \frac{2}{3} (A^\pm)^3 \right] + \cB_\infty(A^\pm).
\end{align}
The level \(\mathrm{k}\) is given by
\begin{equation}\label{eq:kGrelation}
\mathrm{k} = \frac{l}{4G}.
\end{equation}
The spacetime metric is reconstructed from the gauge fields as
\begin{equation}\label{eq:metric}
g_{\mu\nu}=\tfrac{l^2}{2}\Tr\!\big[(A^+ - A^-)_\mu(A^+ - A^-)_\nu\big].
\end{equation}

Using coordinates $(t,r,\theta)$, the gauge fields can be written as
\begin{equation}\label{eq:apmform}
A^\pm=b_\pm^{-1}(d+a^\pm)b_\pm, \qquad 
a^\pm=(\xi_\pm dt \pm p_\pm d\theta)L_0,
\end{equation}
with $p_\pm$ the dynamical fields and $\xi_\pm$ the chemical potentials. where $b^\pm$ are gauge group elements that depend only on the radial coordinate $r$. The flatness condition implies
\begin{equation}\label{eq:maxeq}
\dot p_\pm = \pm \xi'_\pm.
\end{equation}
\( \cdot \) and \( ' \) denote partial derivatives with respect to \( t \) and \( \theta \), respectively.

Requiring a consistent variational principle determines the boundary term and relates the chemical potentials to a boundary Hamiltonian $H^\pm$ {Dutta:2023uxe},
\begin{equation}\label{eq:xidef}
\xi_\pm=-\frac{4\pi}{k}\frac{\delta H^\pm}{\delta p_\pm}, 
\end{equation}
and
\begin{equation}
\cB_\infty^\pm=\pm\int dt\, H^\pm,
\end{equation}
so that the boundary dynamics take the Hamiltonian form
\begin{equation}
\dot p_\pm=\{p_\pm,H^\pm\}_{PB}, 
\end{equation}
and
\begin{equation}\label{eq:Poissonstructure}
\{p_\pm(\theta),p_\pm(\theta')\}_{PB}=\mp \frac{4\pi}{k}\,\partial_\theta\delta(\theta-\theta').
\end{equation}
Thus, the choice of boundary Hamiltonian specifies the dynamics of the fields $p_\pm$, and different Hamiltonians correspond to distinct boundary conditions in $\mathrm{AdS}_{3}$ gravity.

\subsection{A special choice of boundary Hamiltonian}\label{sec:colft}

Following \cite{Dutta:2023uxe}, we choose the boundary Hamiltonian to be the collective field theory Hamiltonian of Jevicki and Sakita,
\begin{equation}\label{eq:ColFTH compact}
H_{\text{ColFT}} = \int d\theta \, \sigma(t,\theta)\!\left[\frac{1}{2}\!\left(\partial_\theta \Pi\right)^2 
+ \frac{\pi^2}{6}\sigma^2(t,\theta) + W(\theta)\right].
\end{equation}
Here \(\sigma(t,\theta)\) and \(\Pi(t,\theta)\) admit a hydrodynamic interpretation as the density and velocity potential of a one-dimensional fluid. The resulting equations of motion take the form of standard fluid equations in the presence of an external potential \(W(\theta)\). Introducing suitable linear combinations of these variables, the dynamics can be reformulated in terms of two fields \(p_\pm\), in which both the Hamiltonian and the equations of motion split into two decoupled sectors. In these variables, the fields \(p_\pm\) satisfy inhomogeneous equations and admit a natural interpretation as the boundaries of the phase space of \(N\) non-relativistic free fermions on a circle~\cite{duttagopakumar,Dutta:2023uxe}. Further details are provided in Appendix~\ref{app:ColFT}.

In this paper, we choose the boundary Hamiltonian, defined in (\ref{eq:xidef}), to be proportional to the collective field theory Hamiltonian written in terms of the variables \(p_\pm\),
\begin{align}\label{eq:bdyHam}
    H^\pm =\frac{\mathrm{k}}{2}H^\pm _{CFT}
    = \pm \frac{\mathrm{k}}{4\pi} \int d\theta \left(  \frac{p_\pm^3}{6}+W(\theta)p_\pm\right).
\end{align}
With this choice, the boundary gauge fields admit a natural interpretation in terms of the phase space of \(N\) non-relativistic fermions, and can equivalently be viewed as encoding the density and velocity fields of a one-dimensional fluid. Within this framework, we restrict our attention to static, non-axisymmetric black flower solutions~\cite{Afshar:2016kjj, Grumiller:2019fmp, Grumiller:2019tyl}.

\section{Black flower solution}\label{sec:black flower solution}

Black flower solutions are static but non-axisymmetric black hole configurations in three-dimensional AdS gravity~\cite{Afshar:2016kjj,Grumiller:2019fmp,Grumiller:2019tyl}. They can be understood as boundary-driven deformations of the BTZ black hole, in which the bulk geometry is completely determined by angularly dependent boundary data while preserving time independence. In contrast to the spherically symmetric BTZ solution, black flower geometries exhibit nontrivial angular modulations of the horizon. These solutions arise naturally in the Chern--Simons formulation of AdS$_3$ gravity and admit a clear interpretation in terms of inhomogeneous boundary fluid configurations.

For time-independent configurations, (\ref{eq:maxeq}) implies that the chemical potentials are constant \cite{Grumiller:2016pqb}. For the boundary Hamiltonian (\ref{eq:bdyHam}), the equations of motion (\ref{eq:ppmeq}) show that the fields \(p_\pm\) acquire a nontrivial \(\theta\)-dependence whenever \(W(\theta)\neq 0\). Consequently, a spherically symmetric BTZ black hole solution is admitted only in the absence of the deformation, i.e.\ for \(W(\theta)=0\), in which case the chemical potentials are constant and the resulting spacetime is static and spherically symmetric \cite{Afshar:2016kjj,Grumiller:2019tyl,Sheikh-Jabbari:2016npa}.

To introduce angular dependence in the boundary fields, one must turn on the potential \(W(\theta)\). Solving the equations of motion then yields boundary fields of the form \(p_\pm(\theta)\sim \sqrt{c_\pm-\lambda W(\theta)}\), while the chemical potentials remain constant, \(\xi_\pm\sim c_\pm\). Although the corresponding bulk metric can be constructed along the same lines as in the homogeneous case, the resulting expressions are technically involved and not especially transparent to present explicitly. 

In this paper, our goal is to construct a black flower solution with a tractable bulk metric, compute its entropy, and explain its microscopic origin through explicit state counting. This is achieved by tailoring the potential term in the boundary Hamiltonian using two parameters \(\lambda_\pm\), chosen so that the resulting geometry remains manageable and allows a clear demonstration of microscopic counting in a spherically non-symmetric setting.

Following the preceding arguments, the boundary Hamiltonian takes the form 
\begin{align}\label{eq:BFHam}
    H^\pm
    = \pm \frac{\mathrm{k}}{4\pi} \int d\theta \left( \frac{p_\pm^3}{6} + \lambda_\pm W(\theta) p_\pm \right).
\end{align}
Although this Hamiltonian can be related to the collective field theory Hamiltonian by a suitable rescaling of the dictionary, in what follows we regard it directly as the defining boundary Hamiltonian that determines the associated bulk geometry.

The equations of motion for the boundary fields \(p_\pm(t,\theta)\) derived from the boundary Hamiltonian are
\begin{equation}\label{eq:peqomfull}
    \dot p_\pm(t,\theta) + p_\pm(t,\theta) p^{\prime}_\pm(t,\theta) + \lambda W^\prime(\theta) = 0 .
\end{equation}
The associated chemical potentials follow from the definition (\ref{eq:xidef}) and are given by
\begin{align}
    \xi^\pm(t,\theta) = \mp \left(\frac{p^2_\pm(t,\theta)}{2} + \lambda_\pm W(\theta) \right).
\end{align}
For time-independent configurations, the chemical potentials must be constant, \(\xi'_\pm(t,\theta)=0\), which implies
\begin{eqnarray}
    p_\pm'(\theta)p_\pm(\theta) + \lambda_\pm W'(\theta) = 0 .
\end{eqnarray}
Solving this equation yields
\begin{eqnarray}
    p_\pm(\theta) = \pm \sqrt{2}\,\sqrt{c_\pm - \lambda_\pm W(\theta)} ,
\end{eqnarray}
where \(c_\pm\) are integration constants.

To obtain a black flower metric in a tractable form, we choose the parameters \(\lambda_\pm\) to be proportional to the constants \(c_\pm\),
\begin{equation}
    \lambda_\pm = \lambda\, c_\pm , \qquad \lambda>0 .
\end{equation}
With this choice, the boundary fields simplify to
\begin{eqnarray}\label{eq:ppmvalueBF}
    p_\pm(\theta) = \pm \sqrt{2\, c_\pm}\,\sqrt{1- \lambda\, W(\theta)} ,
\end{eqnarray}
and the corresponding chemical potentials become constants,
\begin{eqnarray}
    \xi_\pm = \mp c_\pm .
\end{eqnarray}

Next, by redefining the radial coordinate and rescaling the time coordinate according to
\begin{equation}
    \rho = \frac{l}{2} \text{sech}^{-1}\!\left(\frac{2\sqrt{c_+ c_-}\, l^2}{(c_+ + c_-) l^2 - 2 r^2}\right), 
    \qquad 
    t = \gamma^{-1}  \bar t ,
    \qquad 
    \gamma = \frac{l(\sqrt{c_+} + \sqrt{c_-})}{2\sqrt{2}} ,
\end{equation}
the bulk metric can be brought to a BTZ-like form,
\begin{equation}\label{eq:Blackflowermetric}
    ds^2 = - f(r)\, d\bar t^2 + \frac{dr^2}{f(r)} 
    + r^2 \left( \sqrt{1- \lambda\, W(\theta)}\, d\theta 
    + A_{\bar t}(r)\, d\bar t \right)^2 .
\end{equation}
Here the metric functions are
\begin{eqnarray}
    f(r) & = & \frac{r^2}{l^2} - (c_+ + c_-) + \frac{J^2}{r^2} ,
\end{eqnarray}
and
\begin{align}
   A_{\bar t}(r) = \frac{J}{r^2} + \chi ,\quad 
   J = \frac{l}{2}\left(c_- - c_+\right), \quad
   \chi = \frac{\sqrt{c_-}-\sqrt{c_+}}{l(\sqrt{c_-}+\sqrt{c_+})} .
\end{align}
Before proceeding, we examine the structure of the metric in more detail.

 \paragraph{Conditions on \(\lambda\) and \(W(\theta)\):} Since the angular component of the metric behaves as \(g_{\theta\theta} \sim 1-\lambda W(\theta)\), regularity of the geometry requires that the angular coordinate \(\theta\) remain non-degenerate everywhere. This imposes the condition \( 1-\lambda W(\theta)>0 \) for all \(\theta\). Accordingly, constraints must be placed on both the potential \(W(\theta)\) and the parameter \(\lambda\). As a simple example, for the choice \(W(\theta)=\cos\theta\), this condition implies the bound \(|\lambda|<1\). More generally, for any regular periodic function \(W(\theta)\), regularity requires\(|\lambda| < \max_{\theta} |W(\theta)|\).

 \paragraph{Redefining the angular coordinate:} Under these conditions, we may define a new angular coordinate \(\phi\) according to
\begin{equation}
    d\phi = \sqrt{1 - \lambda W(\theta)}\, d\theta \, .
\end{equation}
The periodicity of \(\phi\) is then given by
\begin{eqnarray}
    w(\lambda) = \int_0^{2\pi} \sqrt{1 - \lambda W(\theta)}\, d\theta \, .
\end{eqnarray}
We further introduce a rescaled angular coordinate
\[
\varphi = \frac{2\pi}{w(\lambda)}\, \phi \, ,
\]
so that \(\varphi\) has the standard periodicity \(2\pi\). In terms of these coordinates, the spacetime metric takes the form
\begin{eqnarray}
    \label{eq:Blackflowermetric2}
    ds^2 = - f(r)\, dt^2 + \frac{dr^2}{f(r)} 
    + r^2 \left( \frac{w(\lambda)}{2\pi} d\varphi + \chi\, d\bar t + \frac{J}{r^2} d\bar t \right)^2 .
\end{eqnarray}
Finally, the \(\chi\, d\bar t\) term can be eliminated by the coordinate transformation
\begin{equation}
    d\tilde{\varphi} = d\varphi + \frac{2\pi \chi}{w(\lambda)}\, d\bar t \, .
\end{equation}
In these new coordinates, the metric assumes the standard BTZ form,
\begin{eqnarray}
    ds^2 = - f(r)\, d\bar t^2 + \frac{dr^2}{f(r)} 
    + r^2 \left( \frac{w(\lambda)}{2\pi} d\tilde{\varphi} + \frac{J}{r^2} d\bar t \right)^2 .
\end{eqnarray}

\paragraph{Horizon radii:} The solution admits two horizons \(r_\pm\), determined by the zeros of the lapse function,
\[
f(r_\pm)=0 \, .
\]
These horizon radii can be expressed in terms of the parameters \(c_\pm\) as
\begin{eqnarray}
    r_\pm = \frac{l}{\sqrt{2}}\left(\sqrt{c_+} \pm \sqrt{c_-}\right), 
    \qquad r_+ > r_- \, .
\end{eqnarray}
The extremal limit corresponds to \(c_- = 0\), for which the inner and outer horizons coincide, while the non-rotating solution is obtained when \(c_+ = c_-\).

\paragraph{Euclidean temperature:} To compute the Euclidean temperature of the black hole, we first introduce a co-rotating angular coordinate \(\vartheta\) at the outer horizon \(r_+\), defined by
\begin{equation}
    d\vartheta = d\tilde\varphi + \frac{2\pi}{w(\lambda)} \frac{J}{r_+^2}\, d\bar t \, .
\end{equation}
Near the outer horizon, the lapse function admits the linear expansion
\begin{equation}
    f(r) = \alpha (r-r_+), 
    \qquad 
    \alpha = f'(r_+) = \frac{2(r_+^2 - r_-^2)}{r_+ l^2} \, .
\end{equation}
We then Wick rotate to Euclidean time according to
\begin{eqnarray}
    \bar t = - i \tau \, .
\end{eqnarray}
In these coordinates, the Euclidean metric in the near-horizon region takes the form
\begin{equation}
    ds_E^2 = \alpha (r-r_+)\, d\tau^2 
    + \frac{dr^2}{\alpha (r-r_+)} 
    + r_+^2 \left(\frac{w(\lambda)}{2\pi}\right)^2 d\vartheta^2 \, .
\end{equation}
At fixed \(\vartheta\), the relevant two-dimensional subspace is therefore described by
\begin{equation}
    ds_E^2 \approx \alpha (r-r_+)\, d\tau^2 
    + \frac{dr^2}{\alpha (r-r_+)} \, .
\end{equation}
Requiring regularity of the Euclidean geometry at \(r=r_+\), namely the absence of a conical singularity, fixes the periodicity of the Euclidean time coordinate \(\tau\) to be \(\tau \sim \tau + \beta\), with
\begin{equation}
    \beta = \frac{4\pi}{\alpha} 
    = \frac{2\pi l^2 r_+}{r_+^2 - r_-^2} 
    = \frac{l \pi (\sqrt{c_+} + \sqrt{c_-})}{\sqrt{2}\,\sqrt{c_+ c_-}} \, .
\end{equation}

\paragraph{Periodicity of \(\varphi\):} Thus, smoothness of the Euclidean geometry is imposed while keeping the co-rotating angular coordinate \(\vartheta\) fixed. After Wick rotation, the relation between \(\vartheta\) and the original angular coordinate \(\varphi\) becomes
\begin{equation}
    \vartheta = \varphi - i \frac{2\pi}{w(\lambda)} \left(\frac{J}{r_+^2} + \chi\right)\tau \, .
\end{equation}
Imposing the Euclidean time periodicity \(\tau \sim \tau + \beta\), regularity and single-valuedness of \(\vartheta\) require a compensating shift in \(\varphi\). Consequently, the angular coordinate \(\varphi\) must transform as
\begin{equation}
    \varphi \sim \varphi + i \frac{2\pi}{w(\lambda)} \left(\frac{J}{r_+^2} + \chi\right)\beta \, .
\end{equation}

\paragraph{Triviality of holonomy of the gauge fields:} We now examine the triviality of the holonomy of the gauge fields \(A^\pm\) at the horizon. The holonomy along a closed contour \(\mathcal{C}\) is defined as
\begin{equation}
    \mathbb{H}^\pm_{\mathcal{C}} = \text{Tr}\, \mathcal{P} \exp \left( \oint_{\mathcal{C}} A^\pm \right),
\end{equation}
where \(\mathcal{P}\) denotes path ordering. For any contractible cycle, regularity requires the corresponding holonomy to be trivial.

We choose the contour \(\mathcal{C}\) to lie at fixed \(\vartheta\) and at the outer horizon \(r=r_+\). Along this contour, the holonomy reduces to
\begin{equation}
    \oint_{\mathcal{C}} A^\pm = \oint \left( a_t^\pm\, dt \pm a_\theta^\pm\, d\theta \right).
\end{equation}
After performing the coordinate transformations described above, this expression evaluates to
\begin{eqnarray}
  \oint_{\mathcal{C}} A^\pm =  \pm i\, c_\pm\, \gamma^{-1} \oint d\tau 
    + \sqrt{2 c_\pm}\, \frac{w(\lambda)}{2\pi} \oint d\varphi \, .
\end{eqnarray}
Using the periodicities of the Euclidean time and angular coordinates, \(\tau \sim \tau + \beta\) and the corresponding shift of \(\varphi\), we find
\begin{eqnarray}
  \oint_{\mathcal{C}} A^\pm =  \pm i\, \frac{2\sqrt{2}\sqrt{c_+ c_-}}{l(\sqrt{c_+} + \sqrt{c_-})}\, \beta
    = \pm 2\pi i \, .
\end{eqnarray}
Therefore, the holonomy around the contractible cycle at the horizon is trivial, as required for regularity of the Euclidean black hole solution.

\paragraph{A suitable choice of \(c_\pm\):}
Finally, we parametrise the constants \(c_\pm\) in terms of inverse temperatures \(\beta_\pm\) as
\begin{equation}
    c_\pm = \frac{2\pi^2 l^2}{\beta_\pm^2} \, .
\end{equation}
With this parametrisation, several physically relevant quantities take a particularly simple form. In particular, we obtain
\begin{eqnarray}\label{eq:classicalvalue}
\begin{split}
    \beta & = \frac{\beta_+ + \beta_-}{2}, 
    \qquad 
    r_\pm = \pi l^2 \left(\frac{1}{\beta_+} + \frac{1}{\beta_-} \right),\\
    \xi_\pm & = \mp \frac{2\pi^2 l^2}{\beta_\pm^2} , 
    \qquad 
    p_\pm(\theta) = \pm \frac{2\pi l}{\beta_\pm} \sqrt{1- \lambda W(\theta)} \, .
\end{split}
\end{eqnarray}

\paragraph{Entropy:} The induced metric on the outer horizon, evaluated on a fixed time slice, is given by
\begin{equation}
    ds_H^2 = r_+^2 \left(\frac{w(\lambda)}{2\pi} \right)^2 d\tilde{\varphi}^2 \, .
\end{equation}
The entropy of the black flower solution can then be computed from the area of the outer horizon,
\begin{align}\label{eq:entblackflower}
    S = \frac{A}{4G}
    = \frac{\pi l^2}{4G} \left(\frac{1}{\beta_+} + \frac{1}{\beta_-} \right) w(\lambda) \, .
\end{align}
In the limit \(\lambda \to 0\), one has \(w(\lambda) \to 2\pi\), and the above expression reduces precisely to the Bekenstein--Hawking entropy of the BTZ black hole.

One can obtain the perturbative corrections to the entropy by expanding the effective angular length \(w(\lambda)\) in powers of the deformation parameter \(\lambda\). Retaining terms up to \(\mathcal{O}(\lambda^2)\), the entropy takes the form
\begin{align}\label{eq:entblackflowerlambda3}
S = \frac{A}{4G}
= \frac{\pi^2 l^2}{2G} \left(\frac{1}{\beta_+} + \frac{1}{\beta_-} \right)
\left(1 - \frac{\lambda}{2}\omega_0 - \frac{\lambda^2}{8} \sum_m \omega_m \omega_{-m} \right)
+ \mathcal{O}(\lambda^3) \, .
\end{align}
Here \(\omega_n\) denotes the \(n\)-th Fourier mode of the potential \(W(\theta)\), defined as
\begin{eqnarray}\label{eq:omegan}
\omega_n = \frac{1}{2\pi} \int W(\theta)\, e^{-i n \theta}\, d\theta \, .
\end{eqnarray}

\subsection{Asymptotic charges}

The allowed gauge transformations are those that preserve the asymptotic form of the Chern--Simons connections given in (\ref{eq:apmform}). This requirement restricts the corresponding gauge parameters to lie along the \(L_0\) direction. While such transformations are pure gauge in the bulk, they act non-trivially on the boundary fields \(p_\pm\). In the presence of a boundary, they therefore define genuine asymptotic symmetries with associated canonical surface charges \cite{Banados:1998gg,Banados:1994tn}. These charges form two decoupled and mutually commuting chiral families, are integrable, and exist in infinite number, reflecting the integrable structure underlying the boundary dynamics \cite{Avan:1991ik,Avan:1991kq}.

The boundary-condition--preserving gauge transformations are generated by
\begin{equation}
\lambda_\pm = \eta_\pm(t,\theta)\,L_0 \, ,
\end{equation}
under which the Chern--Simons connections transform as
\begin{equation}
\Delta a^\pm = d\lambda_\pm + [a^\pm,\lambda_\pm] \, .
\end{equation}
The induced variations of the boundary field \(p_\pm\) and the chemical
potential \(\xi_\pm\) are
\begin{equation}
\Delta p_\pm = \pm \partial_\theta \eta_\pm \, , 
\qquad
\Delta \xi_\pm = \partial_t \eta_\pm \, .
\end{equation}
Since the chemical potentials are field-dependent,
\begin{equation}
\xi_\pm = -\frac{4\pi}{k}\,\frac{\delta H_\pm}{\delta p_\pm} \, ,
\end{equation}
consistency of the transformation requires the gauge parameter
\(\eta_\pm(t,\theta)\) to satisfy
\begin{eqnarray}\label{eq:etaeqn}
\begin{split}
\partial_t \eta_\pm(t,\theta)
  =
\mp\frac{4\pi}{k}\,
\Delta \left(\frac{\delta H_\pm}{\delta p_\pm(t,\theta)}\right) 
 = \mp\frac{4\pi}{k}\, \frac{\delta^2 H_\pm}{\delta p^2_\pm(t,\theta)} \partial_\theta \eta_\pm(t,\theta)
 \end{split}
\end{eqnarray}

As shown in \cite{Brown:1986nw,Afshar:2016kjj}, the variation of the canonical charges associated with the asymptotic symmetry
parameters \(\eta_\pm\) is given by
\begin{equation}\label{eq:asympcharges}
\delta Q_\pm[\eta_\pm]
=
\mp \frac{k}{4\pi}
\int d\theta\;
\eta_\pm(t,\theta)\,
\delta p_\pm(t,\theta) \, .
\end{equation}

It turns out that the Hamiltonian itself arises as one of the conserved charges when the gauge transformation parameters are chosen to coincide with the chemical potentials \(\xi_\pm\). This choice is consistent with the condition (\ref{eq:etaeqn}) once the equations of motion (\ref{eq:maxeq}) are imposed. Moreover, one can explicitly verify that the Hamiltonian is conserved on-shell. This particular choice of the gauge parameters \(\eta_\pm\) therefore corresponds to time-translation symmetry of the bulk spacetime metric.

When \(\lambda = 0\), the bulk metric admits an additional Killing vector \(\partial_\theta\), corresponding to rotational symmetry. The conserved charge associated with angular translations is obtained by choosing the gauge parameters as
\[
\eta_\pm = \mp p_\pm(\theta) \, .
\]
For \(\lambda = 0\), this choice is consistent with the condition (\ref{eq:etaeqn}) once the equations of motion are imposed. The resulting conserved charge is identified with the angular momentum of the BTZ black hole geometry and is given by
\begin{eqnarray}\label{eq:Jlambda0}
    \mathcal{J}^{(\lambda=0)}_\pm = \frac{k}{4\pi} \int d\theta\, \frac{p_\pm^2(\theta)}{2} \, .
\end{eqnarray}
Evaluating \(\mathcal{J}_\pm\) on the classical solution \(p_\pm(\theta) = \pm \dfrac{2\pi l}{\beta_\pm}\), one finds that the combination \(\mathcal{J}_+ + \mathcal{J}_-\) correctly reproduces the angular momentum of the BTZ black hole.

In presence of \(W(\theta)\), \(\eta_\pm = \mp p_\pm(\theta)\) fails to satisfy (\ref{eq:etaeqn}) on-shell and hence \(\mathcal{J}_\pm\) is not an asymptotic conserved charge anymore. However, if we treat the parameter \(\lambda\) to be small, then it is possible to compute the corrections to \(\mathcal{J}_\pm\) such that it becomes an asymptotic conserved charge to desired order in \(\lambda\). After a careful calculation we find that
\begin{equation}
    \eta_\pm = \mp \left(p_\pm + \lambda \frac{W(\theta)}{p_\pm(\theta)} - \frac{\lambda^2}{2} \frac{W^2(\theta)}{p_\pm^3(\theta)} \right) + \mathcal{O}(\lambda^3)
\end{equation}
is a consistent choice and the corresponding charge is given by
\begin{eqnarray}\label{eq:Jlambda}
    \mathcal{J}^{(\lambda)}_\pm  = \frac{k}{4\pi} \int d\theta\, \left( \frac{p_\pm^2(\theta)}{2} + \lambda W(\theta) \ln[p_\pm(\theta)] + \frac{\lambda^2}{4} \frac{W^2(\theta)}{p_\pm^2(\theta)} \right) + \mathcal{O}(\lambda^3).
\end{eqnarray}

The classical thermodynamic (high-temperature) limit corresponds to \(l/\beta_\pm \gg 1\). In this regime, the boundary fields scale as \(p_\pm \sim |l/\beta_\pm|\). Consequently, in the classical limit the conserved charges \(\mathcal{J}^{(\lambda)}_\pm\) receive their dominant contributions from the leading term on the right-hand side. Evaluating \(\mathcal{J}_\pm\) on the classical solution (\ref{eq:classicalvalue}), we obtain
\begin{eqnarray}\label{eq:Jpmlambdacorrection}
    \mathcal{J}_\pm 
    = \frac{\pi^2 l^2 c}{6\, \beta_\pm^2} 
    \left[ 1 - \frac{\lambda}{2\pi} \int d\theta\, W(\theta) \right]
    = \frac{\pi^2 l^2 c}{6\, \beta_\pm^2} \left(1 - \lambda \omega_0 \right).
\end{eqnarray}
Thus, for bulk solutions of the form~(\ref{eq:Blackflowermetric}), the asymptotic charges $\mathcal{J}_\pm$ are determined by the relation above. Moreover, the presence of an infinite tower of additional conserved charges~\cite{Dutta:2023uxe} indicates that the system admits an underlying integrable structure.

Our next task is to quantize the system, construct the Hilbert space $\cH$, and determine the degeneracy of states in $\cH$ that reproduce the black flower entropy for a given pair $(\mathcal{J}_+, \mathcal{J}_-)$.

\section{Black flower microstates}

We now turn to the microscopic counting of states associated with a given classical geometry and demonstrate that, in the classical limit, the logarithm of the resulting degeneracy correctly reproduces the entropy (\ref{eq:entblackflower}) of the black flower solution.

To this end, we quantize the boundary theory using the bosonisation prescription developed in \cite{Dutta:2025ypr}, which provides a natural and systematic construction of the boundary Hilbert space. The classical boundary fields \(p_\pm(t,\theta)\), obeying the Poisson algebra (\ref{eq:Poissonstructure}), are promoted to quantum operators and expanded in Fourier modes satisfying a Kac--Moody algebra. Each chiral sector is then bosonized in terms of relativistic Dirac fermions, allowing the boundary fields, as well as composite operators such as \(p_\pm^2\) and \(p_\pm^3\), to be expressed explicitly as fermion bilinears.

This fermionic formulation enables a transparent construction of the conserved charges \(\mathcal{J}_\pm\) and the Hamiltonian operators, both of which are written in terms of bilinear fermionic generators. The resulting Hilbert space is built on the Dirac vacuum, corresponding to a filled Fermi sea, with excited states generated by particle and hole excitations. Of particular interest are the \(\mathbf{n}_\pm\)-particle ground states, which exhibit vanishing quantum dispersion for the operators \(\hat p_\pm\) and therefore serve as semiclassical ground states. More general excitations are naturally organized in terms of Young diagram states, corresponding to irreducible representations of \(U(\infty)\), and provide a convenient basis for counting the microscopic degeneracies relevant for the black flower entropy. Further details of this quantization procedure are presented in Appendix~\ref{app:quantisation}.

In the thermodynamic limit, the bulk solutions~(\ref{eq:Blackflowermetric}) completely fix the asymptotic charges $\mathcal{J}_\pm$. These charges therefore constitute the natural macroscopic data relevant for counting microscopic degeneracies. Our objective is to count all states \(\ket{\psi}\) in the boundary Hilbert space for which the expectation values of the operators \(\hat p_\pm(\theta)\) reproduce the classical black flower profiles (\ref{eq:ppmvalueBF}), and for which the expectation values of the conserved charges \(\hat{\mathcal{J}}_\pm\) agree with their classical values (\ref{eq:Jpmlambdacorrection}), namely
\begin{eqnarray}\label{eq:Jpmclassical}
    \begin{split}
        \bra{\psi}\hat p_\pm(\theta) \ket{\psi} 
        & = \pm \frac{2\pi l}{\beta_\pm}\sqrt{1- \lambda W(\theta)} \, ,\\
        \bra{\psi}\hat{\mathcal{J}}_\pm \ket{\psi} 
        & = \frac{\pi^2 l^2 c}{6\, \beta_\pm^2} \left( 1 - \lambda \omega_0 \right) \, .
    \end{split}
\end{eqnarray}

In the quantum theory, within the bosonisation framework, these conserved quantities are promoted to operators acting on the boundary Hilbert space. In particular, the charges \(\mathcal{J}_\pm\) are represented as
\begin{equation}\label{eq:Joperator}
    \hat{\mathcal{J}}_\pm 
    = \frac{k}{8\pi} \int d\theta \, :\hat p_\pm^2(t,\theta):
    = \frac{1}{c} B^1_0 \, ,
    \qquad 
    B^1_0 = \sum_{m \in \mathbb{Z}} m^K 
    :\psi^\dagger_{m-\tfrac{1}{2}} \psi_{m-\tfrac{1}{2}}: \, .
\end{equation}
Here the normal ordering is taken with respect to the Dirac vacuum, and the operator \(B^1_0\) is expressed explicitly in terms of fermionic bilinears in the bosonized description (see Appendix~\ref{app:quantisation}). Therefore, we have
\begin{eqnarray}
    \langle \hat{\mathcal{J}}_\pm \rangle = \frac{1}{c} \bra{\Psi} B^1_0 \ket{\Psi}.
\end{eqnarray}

In the absence of the potential \(W(\theta)\), it was shown in \cite{Dutta:2025uch} that all bosonic excited states, corresponding to particle--hole pair excitations \(\ket{R^\pm;\mathbf{n}_\pm}\) above the \(\mathbf{n}_\pm\)-particle ground state, account for the degenerate microstates of the BTZ black hole. These excited states are naturally labeled by Young diagrams. The state \(\ket{R^\pm}\) is an eigenstate of the operator \(B^1_0\), with eigenvalue equal to the total number of boxes \(|R^\pm|\) in the corresponding Young diagram \(R^\pm\). 

Consequently, fixing the expectation value of \(\langle \hat{\mathcal{J}}_\pm \rangle\) to its classical value (\ref{eq:Jpmclassical}) amounts to restricting the allowed Young diagram states to those whose total number of boxes matches the classical value of \(\langle \hat{\mathcal{J}}_\pm \rangle\), namely
\[
|R^\pm| = \frac{\pi^2 l^2 c^2}{6 \, \beta_\pm^2} \, .
\]
The logarithm of the resulting degeneracy reproduces precisely the Bekenstein--Hawking entropy of the BTZ black hole\footnote{In principle, one could also perform the microscopic counting by fixing the expectation values of the boundary Hamiltonian in addition to those of \(\hat p_\pm(\theta)\). The boundary Hamiltonian (\ref{eq:BFHam}) can be written in terms of fermionic bilinears and is given by (\ref{eq:bosonisedH0}). The expectation value of the Hamiltonian in a Young diagram state \(\ket{R^\pm}\) is therefore given by
\begin{eqnarray}
    \langle H^\pm \rangle = \frac{\sqrt{3}}{c}\bra{R^\pm} B^2_0 \ket{R^\pm} + \frac{\pi^2 l^2}{\sqrt{3}\beta_\pm^2} \omega_0.
\end{eqnarray}
The matrix element \(\bra{R^\pm} B^2_0 \ket{R^\pm}\) depends on the representation data through the quantity \(\kappa_R\), defined as
\[
\kappa_R = \sum_i l_i (l_i - 2 i + 1),
\]
where \(l_i\) denotes the number of boxes in the \(i\)-th row of the Young diagram \(R^\pm\). Consequently, demanding that \(\langle H^\pm \rangle\) coincide with its classical value amounts to restricting the allowed representations to those with fixed \(\kappa_R\).

However, we find that the constraints imposed by fixing the expectation values of \(\hat p_\pm(\theta)\) and the conserved charges \(\hat{\mathcal{J}}_\pm\) already provide the most restrictive conditions on the allowed states and are therefore sufficient to determine the relevant microscopic degeneracy. Imposing additional constraints, such as requiring both \(\langle \hat{\mathcal{J}}_\pm \rangle\) and \(\langle H^\pm \rangle\) to match their respective classical values, further reduces the number of admissible states by a finite fraction. Nevertheless, such additional restrictions do not affect the leading classical contribution to the entropy.}. Further details of this microscopic counting are presented in Appendix~\ref{app:BTZdegeneracy}.

In the presence of the potential term, our objective is to identify the modified set of degenerate states \(\ket{\psi}\) that satisfy the constraints (\ref{eq:Jpmclassical}). We approach this problem perturbatively in the deformation parameter \(\lambda\), constructing the degenerate states order by order in \(\lambda\). Within this perturbative framework, the states \(\ket{\psi}\) smoothly reduce to the Young diagram states \(\ket{R^\pm}\) in the limit \(\lambda \to 0\), thereby providing a continuous deformation of the BTZ microstates.

%%%%%%%%%%%%%%%%%%%%%%%%%%%%%%%%%%%%%%%%%%%%%%%%%%%%%%%%%%%%%%%%%%%%%%%%%%%%%%%%%%%%%%%%%%%%%%%%%%%%%%%%%%%%%%%%%%%%%%%%%%%%%%%%%%%%%%%%%%%%%%

\subsection{Construction of black flower microstates}

We begin by constructing the perturbatively corrected microstates associated with the black flower geometry. Up to second order in the deformation parameter \(\lambda\), we take the following ansatz for the corrected microstate,
\begin{eqnarray}\label{eq:microstatedef}
    \ket{\Psi^\pm} = \ket{R^\pm} + \lambda ~ \hat g_1\ket{R^\pm} +  \lambda^2 \hat g_2 \ket{R^\pm} + \mathcal{O}(\lambda^3),
\end{eqnarray}
where \(\hat g_1\), \(\hat g_2\) are operators acting on the Hilbert space \(\mathcal{H}\) that encode the perturbative corrections. This expansion represents a systematic deformation of the degenerate BTZ microstates.

We further require the states \(\ket{\Psi^\pm}\) to be normalized to unity. Since the states \(\ket{R^\pm}\) are already normalized, this condition imposes constraints on the operators \(\hat g_1\) and \(\hat g_2\),
\begin{align}\label{eq:g1g2cond}
    \hat g_1^\dagger=-\hat{g}_1,\quad \hat{g}^\dagger_2 + \hat{g}_2 = (\hat{g}_1)^2.
\end{align}

Next, we compute the expectation value of the operator \(\hat p_\pm(\theta)\) in the corrected states and equate it to the classical profile (\ref{eq:ppmvalueBF}). Using the mode expansion of \(\hat p_\pm\) given in (\ref{eq:defptilde}), together with the properties of \(\hat g_1\) and \(\hat g_2\), we obtain
\begin{align}\label{eq:ponmicro}
       \langle p_\pm \rangle = \bra{\Psi^\pm}\hat p_\pm \ket{\Psi^\pm} & = \bra{R^\pm} \hat p_\pm \ket{R^\pm} + \lambda \left(\bra{R^\pm} \hat p_\pm \,\hat g_1 \ket{R^\pm} + \bra{R^\pm}\hat g_1^\dagger \, \hat p_\pm \ket{R^\pm}\right)\nn\\
       & + \lambda^2 \left(\bra{R^\pm} \hat p_\pm \,\hat g_2 \ket{R^\pm} + \bra{R^\pm}\hat g_2^\dagger \, \hat p_\pm \ket{R^\pm} +  \bra{R^\pm}\hat g_1^\dagger \, \hat p_\pm\,\hat g_1 \ket{R^\pm} \right)\nn\\
       &= \frac{2\sqrt{3}}{c}  \mathbf{n}_\pm +\frac{2\sqrt{3}}{c} \lambda  \sum_m \bra{R^\pm} [\alpha_m,\hat g_1] \ket{R^\pm}e^{im\theta} \nn\\
       &+ \frac{2\sqrt{3}}{c}\lambda^2   \sum_m \left(\bra{R^\pm} [\alpha_m,\hat g_2] \ket{R^\pm} - \bra{R^\pm}\hat g_1 [\alpha_m,\hat g_1] \ket{R^\pm}   \right)e^{im\theta}.
\end{align}

To compare with the classical solution, we expand the potential \(W(\theta)\) in Fourier modes (\ref{eq:omegan}) and compute the corresponding classical expression for \(p_\pm(\theta)\),
\begin{eqnarray}\label{eq: pclassical}
\begin{split}
    p_\pm = & \pm \frac{2\pi l }{\beta_\pm} \left( 1-\frac{ \lambda}{2} \omega_0 - \frac{\lambda^2}{8} \sum_{n} \omega_{-n}\omega_n\right)
     \mp \lambda \frac{\pi l}{\beta_\pm} \sum_{m\neq 0} \omega_m e^{i m\theta} \\
     &  \mp \frac{\lambda^2}{4} \frac{\pi l}{\beta_\pm} \sum_{m\neq0} \sum_n \omega_{m-n}\omega_n e^{im\theta} + \mathcal{O}(\lambda^3).
\end{split}
\end{eqnarray}

Matching the zero modes of (\ref{eq:ponmicro}) and (\ref{eq: pclassical}) fixes the corrected particle numbers,
\begin{eqnarray}\label{eq:zeromodecorrection}
    \mathbf n_\pm =  \pm \frac{\pi l c}{\sqrt{3}\beta_\pm}  \left( 1-\frac{ \lambda}{2} \omega_0 - \frac{\lambda^2}{8} \sum_{n} \omega_{-n}\omega_n \right) + \mathcal{O}(\lambda^3).
\end{eqnarray}

Similarly, comparing the coefficients of \(e^{im\theta}\) at order \(\mathcal{O}(\lambda)\) yields
\begin{eqnarray}\label{eq:alpha-g1}
    [\alpha_m, \hat g_1]  = \mp \frac{\pi l c}{2\sqrt{3}\beta_\pm} \omega_m.
\end{eqnarray}
Since this commutator is a \(c\)-number, the operator \(\hat g_1\) must be linear in the oscillators \(\alpha_m\). We therefore adopt the ansatz
\begin{equation}\label{eq:g1form}
    \hat{g}_1=\sum_{m\neq 0} \frac{c_m}{m} \alpha_{-m}.
\end{equation}
which leads to
\begin{equation}\label{eq:cm}
    c_m = \mp\frac{\pi l c}{2\sqrt{3}\beta_\pm}\omega_m.
\end{equation}

Proceeding to order \(\mathcal{O}(\lambda^2)\), we find
\begin{equation}\label{eq:lambda2pval}
[\alpha_m,\hat g_2] = \mp \frac{\pi l c}{2\sqrt{3}\beta_\pm}\omega_m \hat{g}_1 \mp  \frac{\pi l c}{8\sqrt{3}\beta_\pm}\sum_n \omega_{m-n}\omega_n.
\end{equation}
The structure of this commutator implies that \(\hat g_2\) must contain a term proportional to \(\hat g_1^2\) as well as a term linear in \(\alpha_m\). We therefore take
\begin{equation}\label{eq:g2form}
    \hat{g}_2 = \mathcal A (\hat{g}_1)^2 +\sum_{m\neq 0} \frac{\mathcal B_m}{m}\alpha_{-m},
\end{equation}
which yields
\begin{equation}
    \mathcal A=\frac{1}{2},\quad \mathcal B_m = \mp \frac{\pi l c}{8\sqrt{3}\beta_\pm}\sum_n \omega_{m-n}\omega_n.
\end{equation}

This completely fixes the microstates corresponding to the black flower geometry. Since \(\hat g_1\) is linear in \(\alpha_m\), the corrected states \(\hat g_1\ket{R^\pm}\) involve Young diagrams with different numbers of boxes. In the limit \(\lambda\to 0\), all microstates share the same number of boxes, while the \(\lambda\)-deformation lifts this degeneracy. At order \(\mathcal{O}(\lambda)\), the corrected state is a linear superposition of Young diagram states with different box numbers, and this feature persists at higher orders.

Finally, we compute the expectation value of the operator \(\hat{\mathcal J}_\pm\) defined in (\ref{eq:Joperator}). The details of the computation are presented in Appendix~\ref{app:Jcal}, and the result is
\begin{eqnarray}\label{eq:Jhatcorrection}
    \langle \hat{\mathcal J}_\pm \rangle = \frac{1}{c} \left( |R^\pm| + \frac{\mathbf{n_\pm}(\mathbf{n}_\pm+1)}{2} +\lambda^2\,\frac{\pi^2 l^2 c^2}{24}  \sum_{m\neq0}\omega_{-m}\omega_m\right).
\end{eqnarray}
The second term corresponds to a ground-state contribution. Subtracting this piece and comparing with the classical result (\ref{eq:Jpmclassical}), we obtain
\begin{eqnarray}
    |R^\pm| = \frac{\pi^2 l^2 c^2}{6\beta^2_\pm}\left(1- \lambda \omega_0 - \frac{\lambda^2}{4}  \sum_{m\neq0}\omega_{-m}\omega_m \right).
\end{eqnarray}
Thus, we find that turning on spherical deformations not only induces $\lambda$-dependent corrections to the Young diagram states, but also modifies the number of boxes of the leading state $\ket{R^\pm}$.

The resulting entropy is therefore given by the Hardy-Ramanujan asymptotic formula (\ref{eq:Hardy-ramanujan})
\begin{eqnarray}
\begin{split}
    S & = 2\pi \sqrt{\frac{|R^+|}{6}} + 2\pi \sqrt{\frac{|R^-|}{6}} 
    = \frac{\pi^2 l^2}{2G} \left( \frac{1}{\beta_+} + \frac{1}{\beta_-}\right)
    \left(1 - \frac{\lambda}{2} \omega_0-\frac{\lambda^2}{8}  \sum_{m}\omega_{-m}\omega_m \right),
\end{split}
\end{eqnarray}
which precisely matches the gravitational entropy (\ref{eq:entblackflowerlambda3}).

\section{Summary and Outlook}\label{sec:summary}

In this work, we analyzed black flower solutions in three-dimensional AdS gravity within the Chern--Simons formulation, emphasizing the role of boundary Hamiltonians in determining bulk geometry and dynamics. By adopting collective field theory--inspired boundary conditions, we showed that stationary but non-axisymmetric black hole geometries arise naturally from angularly inhomogeneous boundary data, while remaining regular and thermodynamically well defined.

We constructed a tractable class of black flower geometries by introducing a controlled angular potential in the boundary Hamiltonian. This allowed us to explicitly solve the boundary equations of motion, determine the associated bulk metric, and analyze its geometric and thermodynamic properties. The resulting spacetime can be brought to a BTZ-like form with a deformed angular sector, admits regular horizons, and possesses an entropy that depends on an effective angular length. In the limit of vanishing deformation, the geometry and entropy smoothly reduce to those of the BTZ black hole.

In \cite{Afshar:2016kjj}, the entropy of the black flower solution was obtained by exploiting Lifshitz scaling properties and analyzing the asymptotic growth of states in an appropriate limit. The present work adopts a complementary perspective, based on an explicit canonical quantization of the boundary theory. On the microscopic side, we quantized the theory using bosonization, mapping the collective fields to relativistic free fermions. The resulting Hilbert space is naturally organized in terms of Young diagram representations. Fixing the asymptotic charges selects the relevant sector of states, and in the presence of the angular potential the black flower microstates arise as smooth deformations of the BTZ microstates. We showed that the logarithm of the microscopic degeneracy reproduces the gravitational entropy, including its nontrivial dependence on the deformation.

These results demonstrate that black hole entropy in AdS$_3$ continues to admit a precise microscopic interpretation even in the absence of rotational symmetry and in the presence of field-dependent chemical potentials. It would be interesting to explore non-perturbative aspects of the construction, the role of the full tower of conserved charges, and possible extensions to higher-spin or super-symmetric theories as studied in \cite{Banerjee:2025iks}, where similar boundary-driven mechanisms may give rise to new classes of non-axisymmetric black holes.

\paragraph{Acknowledgment} We thank Nabamita Banerjee and Ranveer Singh for useful discussions. We also acknowledge the assistance of AI tools (ChatGPT and Grammarly), which were used solely to improve the grammar and clarity of the manuscript. We are deeply indebted to the people of India for their continued and unconditional support for research in basic science.

\appendix

\section{Collective field theory Hamiltonian}\label{app:ColFT}

The collective field theory equations of motion can be written in terms of $\sigma(t,\theta)$ and \(v(t,\theta) = \partial_\theta \Pi(t,\theta),\). They are given by,
\begin{align}
\partial_t \sigma(t,\theta) + \partial_\theta \bigl(\sigma(t,\theta)\, v(t,\theta)\bigr) = 0\, , \quad 
\partial_t v(t,\theta) + v(t,\theta)\, \partial_\theta v(t,\theta)
= - \partial_\theta \left( \frac{\pi^2}{2}\, \sigma^2(t,\theta) + W(\theta) \right)\,.
\end{align}
The equations of motion admit a natural interpretation in terms of one-dimensional hydrodynamics on the boundary circle \cite{sakita,das-jevicki}. The field \(\sigma(t,\theta)\) represents the fluid density, while \(v(t,\theta)\) plays the role of the local velocity. The first equation is a continuity equation expressing conservation of the boundary density, whereas the second equation is an Euler equation for an inviscid fluid subject to an external potential \(W(\theta)\). The term proportional to \(\sigma^2\) acts as an effective pressure with equation of state \(P(\sigma)\propto\sigma^2\), characteristic of a collective field description of one-dimensional fermions. 
%In the absence of the external potential, the fluid is homogeneous and corresponds to the BTZ black hole, while a nontrivial \(W(\theta)\) induces static but spatially inhomogeneous fluid configurations that give rise to non-axisymmetric bulk geometries such as black flower solutions.
 
To facilitate the analysis, it is convenient to introduce the variables $p_{\pm}(t,\theta)$, 
defined as
\begin{equation}\label{eq:ppmsv-compact}
    p_\pm \;=\; v \pm \pi\sigma
    %\frac{v \pm \pi\sigma}{\kappa_\pm}.
\end{equation}
In terms of these new variables, the fluid equations take the compact form
\begin{equation}\label{eq:ppmeq}
 \partial_t p_\pm +  p_\pm\, \partial_\theta p_\pm 
    + W'(\theta) \;=\; 0.
    %\kappa_\pm \left( \partial_t p_\pm + \kappa_\pm\, p_\pm\, \partial_\theta p_\pm \right) 
    %+ W'(\theta) \;=\; 0.
\end{equation}
The quantities $p_\pm(t,\theta)$ admit a natural interpretation: 
they describe the boundaries of the phase space occupied by 
$N$ non-relativistic free fermions on a circle \cite{duttagopakumar,Dutta:2023uxe}.
%In this formulation, the constants $\kappa_\pm$ appear as scaling parameters. They encode the freedom to rescale the phase space area, thereby fixing the dictionary between the fluid variables and the fermionic phase space variables. We shall exploit this freedom in the next section to construct the black flower solution.

The Hamiltonian then decomposes as
\begin{equation}
H_{\text{ColFT}} = H_{\text{ColFT}}^+ + H_{\text{ColFT}}^-, 
\end{equation}
where,
\begin{equation}\label{eq:ColFTH}
H_{\text{ColFT}}^\pm = \pm \frac{1}{2\pi}\int d\theta 
\left(\frac{p_\pm^3}{6} + W(\theta)p_\pm\right).
%\pm \frac{1}{2\pi}\int d\theta 
%\left( \kappa_\pm\frac{p_\pm^3}{6}+\frac{W(\theta)p_\pm}{ %\kappa_\pm}\right).
\end{equation}
By choosing the boundary Hamiltonian (\ref{eq:xidef}) to be proportional to the collective field theory Hamiltonian, we obtain a direct mapping between bulk geometries and the dynamics of the boundary fluid. Accordingly, we take the boundary Hamiltonian to be given by (\ref{eq:bdyHam}).

\section{Quantization and the Hilbert space}
\label{app:quantisation}

We follow the bosonisation prescription of \cite{Dutta:2025ypr} to construct the Hilbert space. Here we briefly outline the main steps of quantisation.

The boundary fields $p_\pm(t,\theta)$ satisfy the Poisson structure 
(\ref{eq:Poissonstructure}). To construct the boundary Hilbert space, 
we promote this Poisson bracket to a commutator. Expanding the fields 
$p_\pm$ in Fourier modes, we write 
\begin{equation}\label{eq:defptilde}
    p_\pm(t,\theta) \;=\; \frac{2\sqrt{3}}{c} \sum_{n} 
    \alpha^{\pm}_{\pm n}\, e^{i n \theta}.
\end{equation}
The modes $\alpha^\pm_n$ then satisfy the algebra 
\begin{equation}\label{eq:balgebra}
    [\alpha^\pm_m, \alpha^\pm_n] \;=\; m \,\delta_{m+n}.
\end{equation}

Since the two sets of modes form independent copies of the Kac--Moody 
algebra \eqref{eq:balgebra}, we can bosonize them in terms of relativistic 
Dirac fermions $\psi_\pm(t,\theta)$. The correspondence is expressed as\footnote{Similar construction is performed in \cite{Afshar:2016uax, Sheikh-Jabbari:2016npa,Afshar:2017okz}.}
\begin{equation}\label{eq:bosonisation2}
    :\psi^\dagger_\pm(t,\theta)\psi_\pm(t,\theta): 
    \;\equiv\; \frac{c}{4\pi\sqrt{3}}\,p_\pm(t,\theta),
\end{equation}
where the fermions satisfy the equal-time anti-commutators
\begin{eqnarray}\label{eq:psianti}
\begin{split}
   \{\psi_\pm(t,\theta),\psi_\pm(t,\phi)\} &= 0, 
   \quad \{\psi^\dagger_\pm(t,\theta),\psi^\dagger_\pm(t,\phi)\} = 0, \\
   \{\psi_\pm(t,\theta), \psi^\dagger_\pm(t,\phi)\} &= \delta(\theta-\phi).
\end{split}
\end{eqnarray}

The fermionic fields admit the mode expansion (To simplify the presentation, we temporarily suppress the \(\pm\) indices, since both sectors are structurally identical.)
\begin{align}\label{eq:fermionmodes}
    \psi(t,\theta) &= \frac{1}{\sqrt{2\pi}} 
    \sum_{m \in \mathbb{Z}} \psi_{m-\tfrac{1}{2}} \, e^{i m \theta},\nn\\
    \psi^\dagger(t,\theta) &= \frac{1}{\sqrt{2\pi}} 
    \sum_{m \in \mathbb{Z}} \psi^\dagger_{m-\tfrac{1}{2}} \, e^{-i m \theta}.
\end{align}
The corresponding fermionic modes $\psi_q, \psi^\dagger_r$, with 
$q,r \in \mathbb{Z}+\tfrac{1}{2}$, satisfy
\begin{eqnarray}\label{eq:psialgebra}
    \{\psi_q,\psi_r\}=0, \quad 
    \{\psi^\dagger_q,\psi^\dagger_r\}=0, \quad 
    \{\psi_q,\psi^\dagger_r\}=\delta_{q,r}.
\end{eqnarray}

Normal ordering in \eqref{eq:bosonisation2} is defined as
\begin{eqnarray}
    :\psi^\dagger_q \psi_r: \;=\;
    \begin{cases}
        \psi^\dagger_q \psi_r & r>0, \\
        - \psi_r \psi^\dagger_q & r<0 .
    \end{cases}
\end{eqnarray}

Putting this together, the bosonized expression for the boundary fields is\footnote{For additional reviews see \cite{Dhar:1992hr,Dhar:2005qh,Das:1995gd,Das:2004rx,Rao:2000rh,Miranda:2003ga}.}
\begin{equation}
    p_\pm(t,\theta) \;=\; \frac{2\sqrt{3}}{c} 
    \sum_{m,n \in \mathbb{Z}} :\psi^\dagger_{m-n-\tfrac{1}{2}} 
    \psi_{m-\tfrac{1}{2}}:\, e^{i n \theta}.
\end{equation}
Similarly, one finds
\begin{eqnarray}
    :p^2(t,\theta): \;=\; \frac{24}{c^2} 
    \sum_{m,n \in \mathbb{Z}} \left(m - \tfrac{n}{2}\right) 
    :\psi^\dagger_{m-n-\tfrac{1}{2}} \psi_{m-\tfrac{1}{2}}:\, e^{i n \theta}.
\end{eqnarray}

We now express the angular momentum operator \(\hat{\mathcal J}_\pm\) in terms of fermionic operators \cite{Avan:1992gm}
\begin{eqnarray}
    \hat{\mathcal J}_\pm = \frac{\mathrm{k}}{8\pi} \int d\theta :\hat p^2_\pm: = \frac{1}{ c} B^1_0
\end{eqnarray}
where
\begin{equation}
    B^K_n = \sum_{m,n \in \mathbb{Z}} m^K 
    :\psi^\dagger_{m-n-\tfrac{1}{2}} \psi_{m-\tfrac{1}{2}}:
\end{equation}
From (\ref{eq:defptilde}) we see that
\begin{eqnarray}
    B^0_m = \alpha_m.
\end{eqnarray}

The Hamiltonian is given by
\begin{eqnarray}\label{eq:bosonisedH0}
    H^\pm = \frac{\sqrt{3}}{c} B^2_0 + \frac{\pi^2 l^2}{\sqrt{3}\beta_\pm^2} \sum_m \omega_{-m} B^0_m.
\end{eqnarray}

To construct a representation of the algebra (\ref{eq:psialgebra}), we introduce the Dirac vacuum $\ket{0}$, defined by
\begin{eqnarray}
   \psi_{k+\tfrac{1}{2}} \ket{0} &=& 0, \qquad k \geq 0, \\
   \psi^\dagger_{k-\tfrac{1}{2}} \ket{0} &=& 0, \qquad k \leq 0 .
\end{eqnarray}
In other words, the ground state corresponds to a filled ``Dirac sea'' in which all negative-energy modes (with $q<0$) are occupied, while the positive-energy modes (with $q>0$) remain empty.  

Excited states are generated by acting on $\ket{0}$ with creation operators $\psi^\dagger_q$ for $q>0$ and annihilation operators $\psi_q$ for $q<0$. The state $\psi^\dagger_{q>0}\ket{0}$ represents a single fermion excitation above the Dirac sea, whereas $\psi_{q<0}\ket{0}$ corresponds to a hole excitation, describing the removal of a fermion from the sea.

For our purposes, we define the normalized $\mathbf{n}$-particle ground state as
\begin{equation}\label{eq:nstate}
\ket{\mathbf{n}} =
\begin{cases}
   \psi^\dagger_{\mathbf{n}-\tfrac{1}{2}} \cdots 
   \psi^\dagger_{\tfrac{1}{2}} \ket{0}, & \mathbf{n} > 0, \\[6pt]
   \psi_{\mathbf{n}+\tfrac{1}{2}} \cdots 
   \psi_{-\tfrac{1}{2}} \ket{0}, & \mathbf{n} < 0 .
\end{cases}
\end{equation}
States with $\mathbf{n}>0$ correspond to adding $\mathbf{n}$ fermions above the Dirac sea, while states with $\mathbf{n}<0$ describe hole excitations, obtained by removing $|\mathbf{n}|$ fermions from the sea.

One can show that the $\mathbf{n}$-particle ground states satisfy
\begin{equation}
\begin{split}
    \bra{\mathbf{n}_\pm}:{p}_\pm: \ket{\mathbf{n}_\pm} & = \frac{2\sqrt{3}}{c}\mathbf{n}_\pm, \\ \bra{\mathbf{n}_\pm}:{p}^2_\pm: \ket{\mathbf{n}_\pm} & = \frac{12}{c^2} (\textbf{n}_\pm(\textbf{n}_\pm+1)),\\
    \bra{\mathbf{n}_\pm}:{p}^3_\pm: \ket{\mathbf{n}_\pm} & = \frac{12\sqrt{3}}{c^3} (\textbf{n}_\pm(\textbf{n}_\pm+1)(2\textbf{n}_\pm+1)).
    \end{split}
\end{equation}
An interesting property of the $n_{\pm}$-particle ground state is that the quantum dispersion of the operator $\hat{p}_{\pm}$ vanishes identically. We consider these states to be the ground states.

We construct Young diagram states
\begin{align}\label{eq:tmnstate}
    \ket{R^\pm} = \prod_{i} \psi^\dagger_{\mathbf{n}+n_i +\frac12} \psi_{\mathbf{n}- m_i-\frac12}\ket{\mathbf{n}^\pm},\quad n_i, m_i \geq 0.
\end{align}
Each pair \(\psi^\dagger_{\mathbf{n}+n_i +\frac{1}{2}} \psi_{\mathbf{n}-m_i-\frac{1}{2}}\) creates a hole at or below \(k = \mathbf{n}\) and a particle above \(k = \mathbf{n}\). These states can be represented in terms of Young diagrams \(\ket{R}\) (up to an overall sign) of irreducible \(U(\infty)\) representations \cite{Marino:2005sj}. Denoting \(l_i\) and \(v_i\) as the lengths of the \(i^{\text{th}}\) row and column of a diagram, then we have \(m_i = v_i - i\) and \(n_i = l_i - i\).

Further, \(\ket{R^\pm}\) are eigen states of \(B^K_0\) operators. Eigenvalues of \(B^1_0\) and \(B^2_0\) are given by \cite{Dutta:2025uch}
\begin{eqnarray}
    \begin{split}
        \mathcal{E}^1_{R^\pm} & = |R^\pm| + \frac{\mathbf{n_\pm(\mathbf{n}_\pm +1)}}{2}\\
        \mathcal{E}^2_{R^\pm} & = \pm (\kappa_{R^\pm} + (2 |\mathbf{n}_\pm| \pm 1)|R^\pm|) + \frac{\mathbf{n}_\pm(\mathbf{n}_\pm + 1)(2\mathbf{n_\pm}+1)}{6}
    \end{split}
\end{eqnarray}

\section{Microstate for BTZ black hole}\label{app:BTZdegeneracy}

As discussed in \cite{Dutta:2025ypr, Dutta:2025uch} the Young diagram states \(\ket{R^\pm}\) are the corresponding microstates of the BTZ black holes. Because all such states corresponds to
\begin{equation}
    \bra{R^\pm}\hat p^\pm \ket{R^\pm} = \frac{2\sqrt{3}}{c} \mathbf{n^\pm}.
\end{equation}
Comparing the with the classical value of \(p_\pm = \pm \dfrac{2\pi l}{\beta_\pm}\) we have
\begin{eqnarray}
    \mathbf{n}_\pm = \pm \frac{\pi l c}{\sqrt{3}\beta_\pm}.
\end{eqnarray}

The expectation of angular momentum operator (\ref{eq:Joperator}) in the Young diagram state is given by,
\begin{eqnarray}
    \bra{R^\pm} \hat{\mathcal J}_\pm \ket{R^\pm} = \frac{1}{c} \left( |R^\pm| + \frac{1}{2} \mathbf{n}_\pm^2\right)
\end{eqnarray}
Subtracting the ground state contribution and equating the expectation value with the classical value \(\dfrac{\pi^2 l^2 c}{6 \beta_\pm^2}\) given in (\ref{eq:Jpmclassical}) for \(\lambda=0\) we have
\begin{eqnarray}
    |R^\pm | = \frac{\pi^2 l^2 c^2}{6 \beta_\pm^2}.
\end{eqnarray}
As a result, the degeneracy is determined by the number of Young diagrams containing a fixed total number of boxes $|R^\pm|$. In the classical limit, corresponding to large central charge $c$, the number of boxes becomes large. In this regime, the degeneracy admits an asymptotic estimate governed by the Hardy--Ramanujan formula
\begin{equation}\label{eq:Hardy-ramanujan}
    d(p_+, p_-) = e^{\,2\pi \sqrt{\frac{|R^+|}{6}} \;+\; 2\pi \sqrt{\frac{|R^-|}{6}} } \, .
\end{equation}
This gives the statistical entropy to be equal to
\begin{eqnarray}
\begin{split}
S_{\text{stat}}
 = \ln d(p_+, p_-)
& = 2\pi \sqrt{\frac{|R^+|}{6}} + 2\pi \sqrt{\frac{|R^-|}{6}}\\
& = \frac{\pi^{2} \ell^{2}}{2G}
\left( \frac{1}{\beta_+} + \frac{1}{\beta_-} \right)
\end{split}
\end{eqnarray}
which matches the Bekenstein--Hawking entropy of the BTZ black hole.

\section{Calculation of \(\langle\hat{\mathcal{J}}_\pm\rangle\)}\label{app:Jcal}

\begin{eqnarray}
\begin{split}
    \bra{\Psi}B^1_0\ket{\Psi} & = \bra{R^\pm}B^1_0\ket{R^\pm} + \lambda \left( \bra{R^\pm}B^1_0 \, \hat g_1\ket{R^\pm} + \bra{R^\pm}\hat g_1^\dagger\,B^1_0\ket{R^\pm} \right)\\
    & \quad + \lambda^2 (\bra{R^\pm}B^1_0\, \hat g_2\ket{R^\pm} + \bra{R^\pm}\hat g_2^\dagger \,B^1_0 \ket{R^\pm} - \bra{R^\pm}\hat g_1\,B^1_0 \, \hat g_1 \ket{R^\pm}).
    \end{split}
\end{eqnarray}
Since \(\ket{R^\pm}\) is an eigenstate of \(B^1_0\), the $\cO(\lambda)$ term does not contribute. Also using the conditions on \(\hat g_2\) given by (\ref{eq:g1g2cond}) we see that the first two terms in \(\mathcal{O}(\lambda^2)\) is given by
\begin{eqnarray}
     \bra{R^\pm}B^1_0\, \hat g_2\ket{R^\pm} + \bra{R^\pm}\hat g_2^\dagger \,B^1_0 \ket{R^\pm} = \mathcal{E}^1_{R} \bra{R^\pm}\hat g_1^2 \ket{R^\pm}.
\end{eqnarray}
Using the commutation relation,
\begin{equation}
    [B^1_m,B^0_n] =-nB^0_{m+n} + \frac{m(m-1)}{2}\delta_{m+n,0},
\end{equation}
we can simplify the $\cO(\lambda^2)$ term,
\begin{align}
    [B^1_0,\hat g_1]=\sum_{m\neq 0} \frac{c_m}{m} [B^1_0,\alpha_{-m}]= \sum_m c_m \alpha_{-m}.
\end{align}
Further, we have, 
\begin{align}\label{eq:gB1g on R}
    \bra{R_1^\pm}\hat g_1^\dagger\,B^1_0 \,\hat g_1 \ket{R_1^\pm} &= \mathcal{E}^1_R\bra {R^\pm}( \hat{g}_1)^2\ket{R^\pm} + \sum_m c_m\bra {R^\pm} \hat{g}_1 \, \alpha_{-m}\ket{R^\pm},
\end{align}
where \[\mathcal{E}^1_R=|R^\pm| + \frac{\mathbf{n_\pm}(\mathbf{n}_\pm+1)}{2}.\]
We can simplify the last term in \eqref{eq:gB1g on R},
\begin{align}
    \sum_{m\neq0} c_m\bra {R^\pm} \hat{g}_1 \, \alpha_{-m}\ket{R^\pm} = \sum_{m,n\neq0} \frac{c_m c_n}{n}\bra {R^\pm} \alpha_{-n} \, \alpha_{-m}\ket{R^\pm}
\end{align}
This is non-zero only $n+m=0$. Hence we have,
\begin{align}
      \sum_{m\neq0} c_m\bra {R^\pm} \hat{g}_1 \, \alpha_{-m}\ket{R^\pm} &= \sum_n \frac{c_n c_{-n}}{n} \bra {R^\pm} \alpha_{-n} \, \alpha_{n}\ket{R^\pm} = - \sum_{n\neq 0} \frac{c_n c_{-n}}{n} \bra {R^\pm} \alpha_{n} \, \alpha_{-n}\ket{R^\pm}
\end{align}
Hence we have,
\begin{align}
   \sum_{m\neq0} c_m\bra {R^\pm} \hat{g}_1 \, \alpha_{-m}\ket{R^\pm}&= -\frac{1}{2}\sum_n \frac{c_n c_{-n}}{n} \bra {R^\pm}[ \alpha_{n}, \, \alpha_{-n}]\ket{R^\pm} = -\frac{1}{2} \sum_n c_n c_{-n}
\end{align}
Hence we have (\ref{eq:Jhatcorrection}).

\bibliographystyle{hieeetr}
\bibliography{ads}{}

%\bibliography{Bib.bib}
\end{document}

%% file: gdefn.tex
\newcommand{\Tr}{\text{Tr}}

\newcommand{\nn}{\nonumber}

\newcommand{\ben}{\begin{eqnarray}\displaystyle}
\newcommand{\een}{\end{eqnarray}}

\newcommand{\be}{\begin{equation}}
\newcommand{\ee}{\end{equation}}

\newcommand{\bc}{\begin{center}}
\newcommand{\ec}{\end{center}}

\newcommand{\eesp}{\end{split}}
\newcommand{\bsp}{\begin{split}}

\newcommand{\Rmnum}[1]{\expandafter\@slowromancap\romannumeral #1@}
\newcommand{\cB}{\mathcal{B}}

\newcommand{\cH}{\mathcal{H}}

\newcommand{\cL}{\mathcal{L}}

\newcommand{\cO}{\mathcal{O}}

\newcommand{\bensp}{\begin{eqnarray}\begin{split}}
\newcommand{\eensp}{\end{eqnarray}\end{split}}

\newcommand{\bnm}{\begin{matrix}}
\newcommand{\enm}{\end{matrix}}
\def\XXint#1#2#3{{\setbox0=\hbox{$#1{#2#3}{\int}$ }
\vcenter{\hbox{$#2#3$ }}\kern-.6\wd0}}

\renewcommand{\ket}[1]{|#1\big>}
\renewcommand{\bra}[1]{\big<#1|}

\def\cB{\mathcal{B}}